\documentclass{emulateapj}

\usepackage{graphicx,rotating}

\newcommand{\rxte}{\textit{RXTE}}
\newcommand{\inte}{\textit{INTEGRAL}}

\def\grs{GRS~1915$+$105}
\def\h17{H1743$-$322}

\shorttitle{Characterizing the radio -- X-ray connection in GRS 1915+105}
\shortauthors{Prat et al.}

\begin{document}

\title{Characterizing the radio -- X-ray connection in GRS 1915+105 }


\author{L. Prat and J. Rodriguez}
\affil{Laboratoire AIM, CEA/IRFU - CNRS/INSU - Universit\'e Paris Diderot,}
\affil{CEA DSM/IRFU/SAp, Centre de Saclay, F-91191 Gif-sur-Yvette, France}
\email{lionel.prat@cea.fr}

\author{G. G. Pooley}
\affil{Astrophysics, Cavendish Laboratory, University of Cambridge CB3 0HE, UK}


\begin{abstract}

We analyzed radio and X-ray observations of \grs, between May 1995 and June 2006, focusing on the times characterized by radio flares and cycles of hard dips -- soft spikes in the X-ray lightcurve. Assuming these flares to be discrete ejections, we applied a plasmon model to the radio data, with good agreement with the lightcurves. We fitted a total of 687 radio flares with a standard model of a plasmon. We found that the distribution of width is $t_0=1160$ s with an rms deviation of 360 s, while that of the amplitude is $S_{max}=59$ mJy with an rms deviation of 28 mJy. The distribution of width is thus rather peaked, while that of the amplitude not.

Regarding radio and X-ray links, this study confirms previous observations on smaller datasets, namely that X-ray cycles of hard dips -- soft spikes are always followed by radio flares. A strong correlation is found between the length of X-ray ``dips" in the X-ray lightcurves, and the amplitude and fluence of the subsequent radio oscillations. A model of an exponential rise of the form $L_{\mathrm{15}Ê\textrm{ } \mathrm{GHz}}$($\Delta$t) $= L_{max} (1-\exp(-(\Delta t-\Delta t_{min})/\tau)$ is in good agreement with the observations, with the maximum fluence $L_{max}$ of the order 70 Jy.s, and the characteristic time $\tau$ of the order 200-500 s. We discuss possible physical interpretations of this correlation, regarding the nature of the ejected material, and the physical process responsible for the ejection.

\end{abstract}



\section{Introduction}

One of the main issues of Black-Hole Binary (BHBs) physics is the link between accretion and ejection processes. Indeed, if the strong gravitational potential favors naturally the accretion of matter, the occurrence of relativistic ejections of plasma in these systems is still far from being understood. A very promising object to conduct this study is \grs.

Indeed, \grs\ is a unique source in many respects. On the one hand, its main features classify it as a microquasar. \grs\ is a low-mass X-ray binary (LMXB), composed of a K-type star orbiting a Black Hole (BH) of $14.0 \pm 4.4 M_{\sun}$ \citep{Harlaftis:2004}. As in most LMXBs, its X-ray spectrum can be described with a soft disc blackbody (with kT $\sim$ 1--2 keV), and a hard power law extending to $\ge$200 keV \citep[e.g.][]{Rodriguez:2008b}. These are interpreted as an accretion disc and a Comptonizing region (often called ``corona"), respectively. At times, it displays superluminal ejections with true bulk velocities $\ge$0.9c \citep{Mirabel:1994}, or a steady compact jet \citep{Dhawan:2000,Fuchs:2003}. This overall description corresponds to the ``canonical" microquasar.

\begin{figure*}
	\center
	\includegraphics[width=0.7\textwidth]{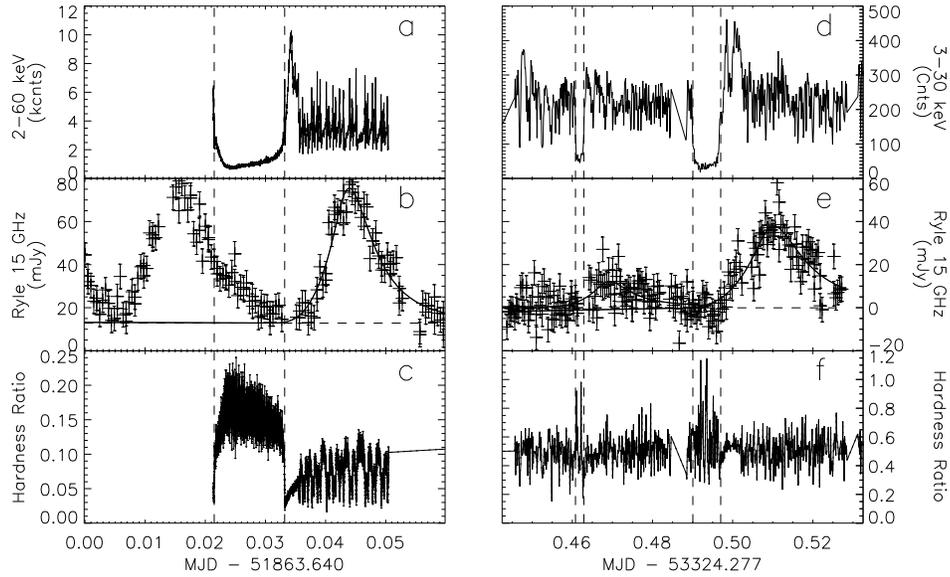}
	\caption{Simultaneous X-ray and Ryle observations of \grs. On MJD 51863.64, \grs\ was in the variability class $\nu$ (left), on MJD 53324.7 it was in class $\lambda$ (right). The plots show on the top the 2--60 keV \rxte\ lightcurve (panel (a)), and the 3--30 keV JEM-X lightcurve (panel (d)). Panels (b) and (e) display the 15 GHz radio lightcurve, while panels (c) and (f) display the hardness ratios. The vertical dotted lines mark the beginning and the end of the X-ray dips on each panel. The horizontal dotted lines are the estimated level of background radio emission. On top of these lines, the continuous lines show the fit of the plasmon model on the radio data. Both observations correspond to a confidence index of 1 (see section \ref{confi}).}
	\label{exemples1}
\end{figure*}

\begin{figure*}
	\center
	\includegraphics[width=0.7\textwidth]{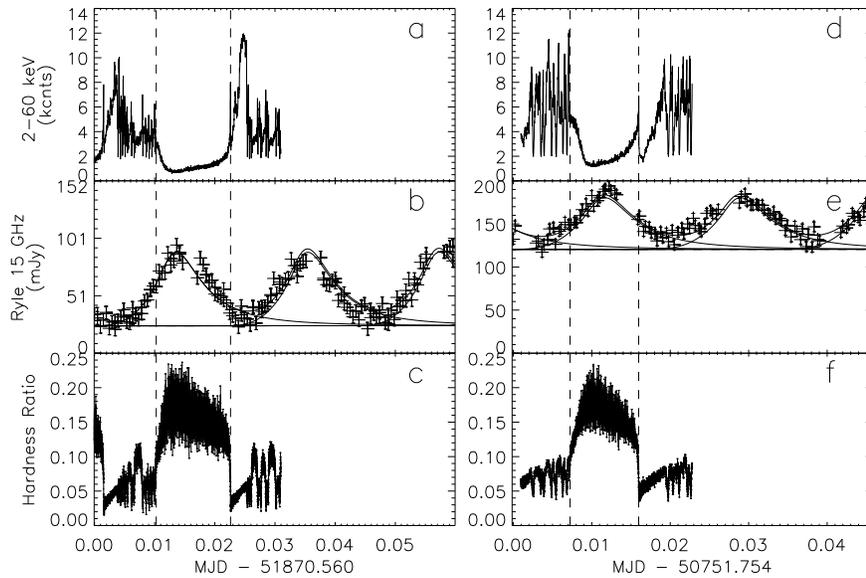}
	\caption{Simultaneous \rxte\ and Ryle observations of \grs, during variability class $\beta$. Panels and lines are identical to Fig. \ref{exemples1}. On the left side, the radio lightcurve corresponds to a confidence index of 2, while on the right side it corresponds to an index of 3 (see section \ref{confi}).}
	\label{exemples2}
\end{figure*}

On the other hand, \grs\ shows a unique wealth of behavior \citep[see, e.g., the review by][]{Fender:2004}. Contrary to transient microquasars, which are usually active for a few months between quiescent states, \grs\ has been constantly active for the 17 years since its discovery with \textsl{GRANAT} \citep{CastroTirado:1992}. It shows a very high level of variability, which may be linked to the fact that \grs\ has a very high accretion rate compared with most other LMXBs. \citet{Belloni:2000} classified its variability into 12 separate classes, which have been observed to recur almost identically over the years. In particular, the X-ray lightcurves display cycles of hard dips -- soft spikes specific to \grs. They are characterized by phases of low X-ray luminosity and hard spectra, which last between $\sim$10 and $\sim$2000s (see Fig. \ref{exemples1} and \ref{exemples2} for sample lightcurves). These phases are ended by a short spike ($\sim$10s), which marks the end of the hard phase. Note that because of the scaling, these spikes, although present, are not distinguishable on Fig. \ref{exemples1}, right. Such features (hard X-ray dip ended by a spike) will be labeled simply ``cycle" in the following. After the spike, the spectrum of the source becomes softer, and the X-ray luminosity decreases for a while. Usually, the light curve then starts to increase again and shows a second spike, longer ($\sim$100s), softer and brighter than the one marking the end of the hard dip. This second spike marks the beginning of a soft and highly variable phase, which ends when a new hard dip begins.

The 12 \grs\ classes can, in turn, be interpreted as transitions between three basic states, labeled A,B and C. Using the classification of \citet{Homan:2005}, these 3 states can be related to the canonical Soft State, Soft Intermediate State and Hard Intermediate State, respectively. Cycles can thus be described as short C states separated by sequences of A and B states.

\clearpage

In the radio range, \grs\ displays also a very high level of variability, which is related to the 12 variability classes. In particular, strong radio oscillations or isolated flares are detected, which are clearly related to the X-ray dips \citep{Pooley:1997,Mirabel:1998}. The power required to generate the synchrotron emission of these oscillations is likely to be a very significant fraction of the entire accretion energy of the system \citep{Fender:2000,Meier:2001}. Radio oscillations are particularly strong during $\beta$ and $\nu$ classes \citep{KleinWolt:2002}. The $\alpha$, $\lambda$, $\theta$, $\kappa$ and $\rho$ classes are characterized by a weak or variable radio flux. 

Since X-ray cycles and radio oscillations seem to occur at the same times, it is tempting to look for a link between them. Previous analyses \citep{Pooley:1997,KleinWolt:2002} have indeed found such a link, namely that X-ray cycles are followed by radio flares during $\beta$ and $\nu$ classes. \citet{Rodriguez:2008a} showed that this behavior can be extended to $\lambda$ classes, and that the time lag between the X-ray spike at the end of a cycle and the maximum of the subsequent radio flare is roughly constant, regardless of the characteristics of the preceding dip. This led to the interpretation that the spike is the trigger of the radio flare. 

This possible X-ray -- radio link, which is specific to \grs, can be related to the behavior of ``conventional" transients. Indeed, the spike also marks an abrupt transition from harder to softer spectral states, and is followed by radio activity. In the temporal domain, timing analyses suggest that this transition is associated to type B quasi-periodic oscillations \citep[QPOs,][]{Soleri:2008}. These sequence of events and timing behavior are in analogy with the detection of major radio flares and type B QPOs at the time of X-ray spectral transitions in many microquasars \citep{Fender:2009}.

The purpose of this paper is to explore the possible link between X-ray dips and radio flares, with an extended data set and a physical treatment of the radio ejections. Our analysis is centered on $\alpha$, $\beta$, $\lambda$ and $\nu$ classes, which are well covered by observations. A few words will be said about the other cyclic classes in section \ref{other}.

Section 2 starts with a description of the available data, and the reduction process. Section 3 describes the method and models used to analyze the radio data, while section 4 explores the links between radio and X-ray behavior. In section 5, we discuss the physical interpretation of these links.

\section{Observations and Data reduction}
In the radio range, the Ryle Telescope (RT) provides an extensive coverage of \grs\ between May 1995 and June 2006. Its sensitivity (a few mJy with a 32s temporal binning) is sufficient to follow discrete ejection events in \grs\ and adjust physical models to represent them.

\subsection{Radio data: Ryle Telescope}
RT observations follow the scheme described by \citet{Pooley:1997}. The setup measured the Stokes parameter $I+Q$, calibrated using the phase calibrator B~1920+154, and the amplitude calibrators 3C~48 or 3C~286. The resulting 15.2 GHz light-curves were binned at 32s. Using data taken during a quiet period, we estimated a standard deviation of 5.7 mJy, and used this value as nominal error bars.

\subsection{X-ray data: \rxte/PCA and \inte/JEM-X}
The \rxte\ data were reduced with the {\tt{HEASOFT}} v6.5 software package, as described in \citet{Rodriguez:2008a}. We extracted 1s resolution light curves from the Proportionnal Counter Array (PCA) in the 2--60 keV range, as well as in the two energy bands defined in \citet{Belloni:2000}: 2--5.7 keV and 5.7--14.8 keV. We then computed the Hardness Ratio HR=5.7--14.8/2--5.7 keV.

We also used data from the \inte\ monitoring campaign described in \citet{Rodriguez:2008a}. The \inte\ data were reduced using the standard {\tt{Off-line Scientific Analysis (OSA)}} v7.0 software package \citep{Goldwurm:2003}. Data from the JEM-X instruments were reduced to construct 10s lightcurves in the 3--7, 7--15 and 3--30 keV energy range (channels 46--95, 95--159 and 46--210 respectively). Since JEM-X is less sensitive than the PCA, the Hardness Ratio (HR) is too noisy to be more than a hint to the classification of variability classes.

\section{A look at the radio flares}

\begin{figure}
	\plotone{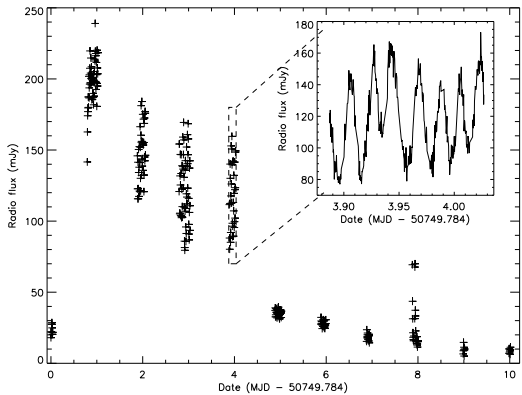}
	\caption{Radio lightcurve of the MJD 50749--50759 major flare seen by the Ryle Telescope. The insert shows a zoom on the lightcurve, where smaller flares are clearly visible on top of the major event.}
	\label{Bigpic}
\end{figure}

In the radio range, \grs\ displays a very rich behavior, characterized by the presence of discrete events. Two types of flares are visible. Firstly, a few giant flares have been detected, which last several days and reach hundreds of mJy (Fig \ref{Bigpic}). Then, numerous flares, with a typical amplitude of a few tens of mJy are also visible, which last around 1 hour. These smaller flares are either isolated or repetitive, and several successive radio flares can overlap in the radio lightcurve (Fig. \ref{exemples2}). Before looking at X-ray observations and variability classes, we tried to characterized these small flares, i.e. their width, amplitude and recurrence period.

To do this, we searched for radio oscillations in the 11 years of RT data. They contain about $\sim$1000 clearly-identified flares. In order to study these flares, we need an emission model to map their evolution.

\subsection{Flare models}
Multiwavelength studies provided good confidence that synchrotron processes are responsible for the emission during radio flares. However, doubts remain on the exact nature of the synchrotron emitting medium, and on its geometry. The most widely used model was described by \citet{vanderLaan:1966}, who calculated synchrotron emission from an adiabatically expanding cloud. This model was first developed for Active Galactic Nuclei (AGNs), for which it is still commonly used. In the case of \grs, \citet{Mirabel:1998} observed delays in the lightcurves at different radio and infrared wavelengths compatible with this plasmon model.

Another possibility was described by \citet{Hjellming:1988}. Starting with the van der Laan model, they calculated the emission from conical twin jets, and adequately predicted the emission of several X-ray binaries. They also provided a geometrical refinement for the van der Laan model, leading to slightly different radio lightcurves for expanding spherical bubbles. 

\citet{Kaiser:2000} described an internal shock model for the observed synchrotron emission. In this scheme, a quasi-continuous jet emission is ``lit-up" by shock fronts traveling along the jets. This model led to similar results as the plasmon model, although with a smaller power emitted by the central engine.

\medskip
Up to now, no scenario is clearly favored by the observations. In this paper, we are first interested in the morphological parameters of the flares, i.e. amplitude, width and fluence. For these parameters, every model will produce approximately the same values, as long as it describes accurately the observations. Therefore, we used the simplest one in our study, i.e. the van der Laan plasmon model, but it should be noted that the physical nature of the flares could be different from a simple spherical plasmon.

\subsection{Description of the model adopted}

In the \citet{vanderLaan:1966} model, flaring emission is produced through the adiabatic expansion of an initially optically thick blob of synchrotron-emitting relativistic electrons. During the first phase, the increase in the blob's surface area causes an increase in the flux. Then, the curve turns over as the plasma becomes optically thin because of the reduction in magnetic field, the adiabatic cooling of the electrons, and the reduced column density as the blob expands. Simultaneous observation at a lower frequency will show the same fractional rate of increase, but the maximum will be reached later and have a smaller value.

As a starting point, we assume that the electrons have an energy spectrum $n(E) \propto E^{-p}$. The synchrotron optical depth at frequency $\nu$ of an homogeneous and spherical blob then scales as

$$
\tau_\nu (t) = \tau_0 \left( \frac{\nu}{\nu_0} \right) ^{-(p+4)/2} \left( \frac{R(t)}{R_0}Ê\right) ^{-(2p+3)}
$$

where $R(t)$ is the radius of the spherical blob at a given time. The flux density then scales as

$$
S_\nu(R) = S_0 \left( \frac{\nu}{\nu_0} \right) ^{5/2} \left( \frac{R}{R_0}Ê\right) ^3 \frac{1-\textrm{exp}(-\tau)}{1-\textrm{exp}(-\tau_0)}
$$

Here, $R_0$, $S_0$ and $\tau_0$ are the size, flux density and optical depth at the peak frequency of the synchrotron spectrum $\nu_0$ \citep{vanderLaan:1966}. $\tau_0$ only depends on $p$ through the condition

$$
e^{\tau_0} - \tau_0 (p+4)/5 -1 = 0
$$

Thus, given the particle energy spectral index $p$ and the peak flux $S_0$ at a given frequency $\nu_0$, this model predicts the variation in flux density at any other frequency as a function of the expansion factor $(R/R_0)$. A model for $R(t)$ is needed to map this relation to time: we will assume a simple linear expansion at constant speed $v_{exp}$. The major ejections detected from \grs\ indicate an expanding velocity of $\sim$ 0.8c \citep{Fender:1999} This model predicts a sharp increase of luminosity, followed by an exponentially decreasing tail. The last parameter, the energy spectral index $p$ depends on the source at hand, and is usually taken between 1 and 3. 

In order to provide precise measurements of the plasmon parameters, the model requires simultaneous monitoring at several frequencies. In particular, the peak frequency $\nu_0$ cannot be determined without a spectral monitoring. Since we observe at a fixed frequency of $\nu=15.2$ GHz only, we will use a simplified version of the formula. We define the characteristic length of a flare to be
$$
t_0=\frac{R_0}{v_{exp}} \tau_0^{\frac{1}{(2p+3)}} \left( \frac{\nu_0}{\nu} \right) ^{\frac{(p+4)}{(4p+6)}}
$$

This leads to the final fitting function

$$
S_\nu(t) = S_{max} \left( \frac{t}{t_0}Ê\right) ^3 \left(1-\textrm{e}^{- \left( \frac{t}{t_0} \right) ^{-(2p+3)}} \right)
$$

Thus, the model will have 3 free parameters: the characteristic length of the flare $t_0$, the maximum amplitude $S_{max}$ and the initial time of the ejection.

\medskip
Several radio observations show a more complex behavior: distinct radio oscillations are visible on top of a strong, slowly varying ``background flux". Fig. \ref{Bigpic} shows an example of such observations: a very bright flare occurs between MJD 50749 and 50759, on top of which smaller flares are clearly distinguishable. In order to isolate the contribution from individual flares, one thus has to remove the contribution from the background flux. To do this, we adopted an empirical solution: since the additional flux varies on a timescale larger than a few radio oscillations (see Fig. \ref{exemples2}), we assumed the background flux to be linear over a few flares. Thus, we removed from the radio flux a linear function of the form $S_{back}(t) = At + K_{0}$, where $A$ is computed by a linear fit to the radio data, and $K_0$ is let free under the condition that $S_{back}$ remains below the minima of the radio lightcurve. Under this definition, $S_{back}(t)$ represents the source flux underlying the small flares: a fraction of the flare flux will be incorporated in this quantity, which will lead to a slight underestimation of the flare fluxes.

\subsection{Typical radio flare parameters}
Once applied to the observations, the plasmon model adequately fits all data, with a reduced chi-squared $\chi_\nu^2$ between 1 and 3. This is especially true for isolated flares, when no strong background correction is needed. Fig. \ref{exemples1}, panels (b) and (e), shows such fits, with $\chi_\nu^2 = 1.26$ and $1.79$ for the two flares peaking on MJD 51863.68 and MJD 53324.79, respectively. These good results give us good confidence in our approach.

We checked for the influence of the spectral energy index $p$ of the electrons for the predicted lightcurve. $p=2$ and $p=3$ gives very similar results: the shape of the predicted flare is similar, and adequately fits the observations. Using $p=1$, however, the predicted flare is narrower, and fails to reproduce the observations. In the following, the results are given for the nominal value $p=2$.

\begin{figure}
	\plotone{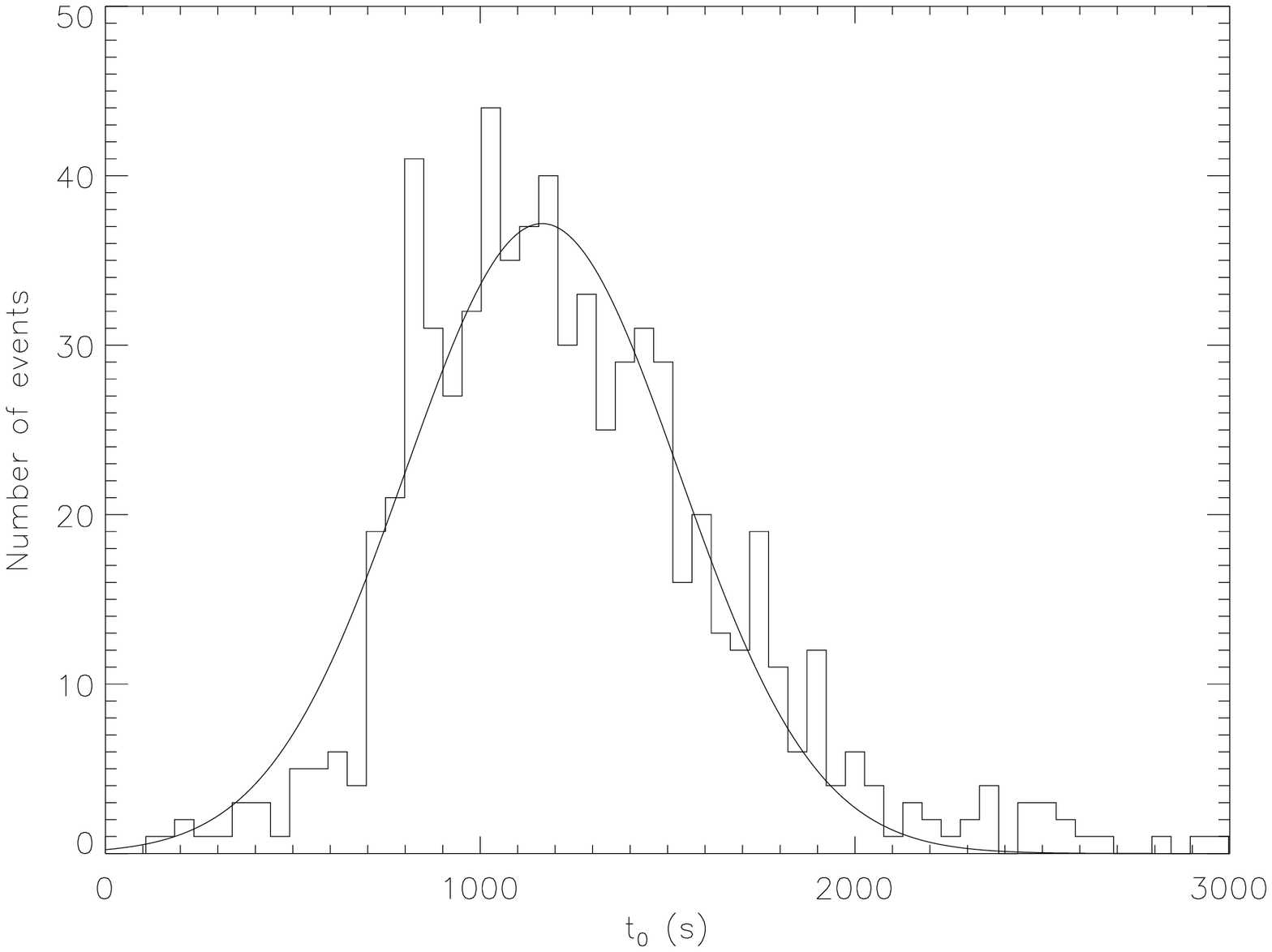}
	\plotone{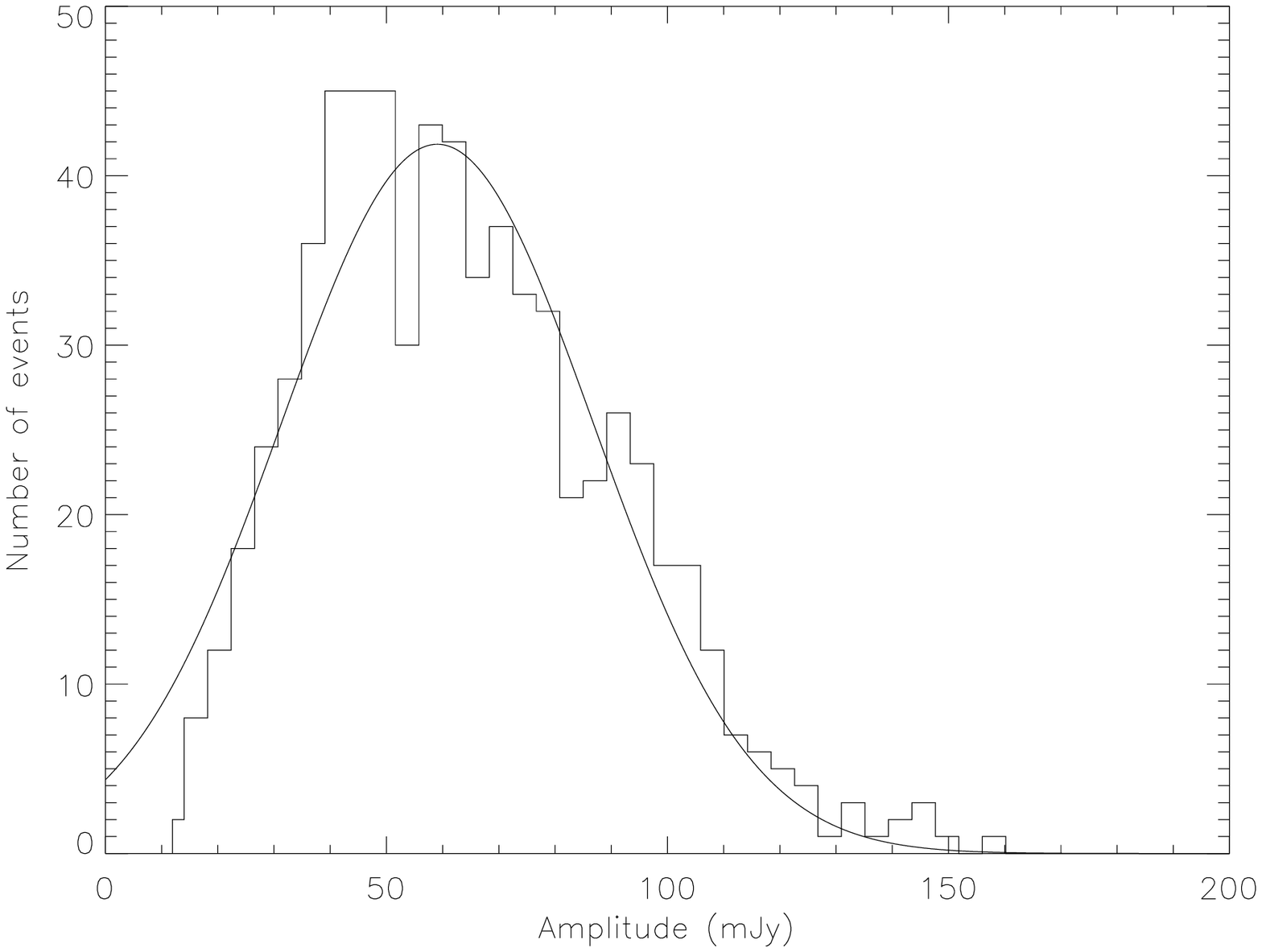}
	\caption{Distribution of the characteristic width and amplitude of the radio flares. The histograms are computed using 687 flares. The continuous lines are gaussian fits to the histograms.}
	\label{histos}
\end{figure}

\medskip
In order to extract the main characteristics of the flares, we applied the plasmon model systematically to the RT data. In doing so, we excluded very noisy data or incomplete flares. A total of 687 flares have been fitted. Fig. \ref{histos} presents the distribution of their characteristic width (given by the $t_0$ parameter) and amplitude. $t_0$ varies between 0 and 3000 s, while the maximum amplitude varies between 10 and 150 mJy.

The distribution of flare width and amplitude appear to be gaussian. Gaussian fits to these histograms give $t_0=1160$ s with an rms deviation of 360 s, and a mean amplitude of $S_{max}=59$ mJy with an rms deviation of 28 mJy.

Note that the distribution of these parameters is biased by four main uncertainties. Firstly, very small flares will hardly be distinguishable from the background noise, which varies between 5 and 6 mJy on the 32s bin lightcurve. Thus, the histograms are truncated at small values. Then, the uncertainty in the background correction will tend to increase the width of the distribution: the linear model is a strong approximation. 

In the case of successive flares, an overlap between them can occur. This was first not taken into account in the background correction. The uncertainty this introduces will tend to shift the maximum of the $t_0$ and $S_{max}$ distributions towards higher values. The effects of overlapping flares are considered in the next section. Finally, in the few cases where, because of the simplified nature of the model, the fit leads to a rather high $\chi_\nu^2$ around 3, the flare parameters are not well constrained. In this case, the effect on the distributions is an increase of their widths (rms deviations).

\section{Radio -- X-ray links}

\subsection{Relevant observations}
We cross-correlated the log of the radio pointings to X-ray observatories, and found a total of 352 \rxte\ pointings and 7 \inte\ observations simultaneous with RT data.

Since our purpose is to study the radio behavior during X-ray cycles, we again filtered our dataset to retain only those observations with good simultaneous coverage. We looked for observations with (1) enough X-ray coverage to detect clearly a hard X-ray dip and (2) radio coverage during the following hour to detect a possible activity. A total of 28 \rxte\ pointings and 3 JEM-X observations display these characteristics for $\alpha$, $\beta$, $\lambda$ and $\nu$ classes. These pointings contain 54 hard X-ray dips with radio coverage, which are listed in table \ref{logg}.

\begin{deluxetable}{ccccc}
\tabletypesize{\small}
\tablewidth{0pt}
\tablecaption{List of observations with simultaneous X-ray and radio coverage, during classes $\alpha$, $\beta$, $\lambda$ and $\nu$.\label{logg}}
\tablehead{\colhead{Start Date} & \colhead{Class}  & \colhead{X-ray} &
\colhead{Dip length} & \colhead{Confidence}\\
\colhead{(MJD)} & \colhead{} & \colhead{instrument} & \colhead{(s)} & \colhead{Index}
}

\startdata
50381.495 & $\nu$ & PCA &            1526$\pm$3 \phn & 2 \\
                  & $\nu$ & PCA &            1333$\pm$5 \phn & 2 \\
                  & $\nu$ & PCA &            $>$ 1800 & 2 \\
50698.658 & $\beta$ & PCA & \phn 562$\pm$5 \phn & 1 \\
                  & $\beta$ & PCA & \phn 554$\pm$1 \phn & 1 \\
                  & $\beta$ & PCA & \phn 542$\pm$1 \phn & 1 \\
50751.687 & $\beta$ & PCA & \phn 767$\pm$1 \phn & 3 \\
                  & $\beta$ & PCA & $>$ 515 & 3 \\
50751.754 & $\beta$ & PCA & \phn 766$\pm$1 \phn & 3 \\
51343.042 & $\beta$ & PCA & $>$ 410 & 3 \\
                  & $\beta$ & PCA &        1000$\pm$2 \phn & 3 \\
                  & $\beta$ & PCA & $>$ 440 & 3 \\
                  & $\beta$ & PCA & \phn 884$\pm$2 \phn & 3 \\
51352.957 & $\beta$ & PCA & $>$ 1430 & -- \\
                  & $\beta$ & PCA &         1411$\pm$2 \phn & -- \\
51353.025 & $\beta$ & PCA & $>$ 1400 & 3 \\
51476.639 & $\beta$ & PCA & \phn 756$\pm$3 \phn & 3 \\
                  & $\beta$ & PCA & $>$ 400  & 3 \\
51814.727 & $\nu$ & PCA & $>$ 1830 & 3 \\
51863.594 & $\nu$ & PCA &            1125$\pm$1 \phn & 1 \\
51863.661 & $\nu$ & PCA &            1009$\pm$1 \phn & 1 \\
51870.554 & $\beta$ & PCA & $>$ 540 & 2 \\
                  & $\beta$ & PCA &         1076$\pm$2 \phn & 2 \\
51870.627 & $\beta$ & PCA & \phn 948$\pm$2 \phn & 3 \\
                  & $\beta$ & PCA & $>$ 530 & 3 \\
                  & $\beta$ & PCA & \phn 955$\pm$2 \phn & 3 \\
                  & $\beta$ & PCA & $>$ 740 &3 \\
51877.522 & $\nu$ & PCA &           1470$\pm$10 & -- \\
                  & $\nu$ & PCA &           $>$ 700  & -- \\
51877.599 & $\nu$ & PCA &           1828$\pm$10 & 2 \\
                  & $\nu$ & PCA &  $>$ 1400 & 2 \\
51885.509 & $\beta$ & PCA & \phn 925$\pm$75 & 3 \\
51954.302 & $\nu$ & PCA &            1140$\pm$5 \phn & 2 \\
51954.441 & $\nu$ & PCA &  $>$ 815 & 1 \\
52108.806 & $\nu$ & PCA &  $>$ 1800 & 2 \\
52108.877 & $\nu$ & PCA &  $>$ 1400 & 2 \\
52171.799 & $\lambda$ & PCA & \phn 140$\pm$2 \phn & 1 \\
                  & $\lambda$ & PCA & \phn\phn 72$\pm$2 \phn & 1 \\
                  & $\lambda$ & PCA & \phn\phn 82$\pm$2 \phn & 1 \\
52500.893 & $\alpha$ & PCA &      1975$\pm$20 & -- \\
53150.057 & $\beta$ & PCA & \phn 857$\pm$3 \phn & 3 \\
53296.373 & $\nu$ & JEM-X &        1826$\pm$20 & 1 \\
                  & $\nu$ & JEM-X &        1727$\pm$20 & 1 \\
53296.728 & $\nu$ & PCA &  $>$ 1400 & 1 \\
53296.794 & $\nu$ & PCA &  $>$ 1000 & 1 \\
53318.559 & $\beta$ & PCA & $>$ 200 & 1 \\
                  & $\beta$ & PCA & \phn  794$\pm$2 \phn & 1 \\
53324.277 & $\lambda$ & JEM-X & \phn 130$\pm$20 & 1 \\
                  & $\lambda$ & JEM-X & \phn 180$\pm$10 & 1 \\
                  & $\lambda$ & JEM-X & \phn 600$\pm$20 & 1 \\
53503.633 & $\beta$ & JEM-X & \phn     420$\pm$20 & 2 \\
                  & $\beta$ & JEM-X &\phn      460$\pm$10 & 2 \\
53703.573 & $\nu$ & PCA & $>$ 700 & 2 \\
53703.639 & $\nu$ & PCA & \phn            827$\pm$10 & 2 \\

\enddata
\end{deluxetable}

Fig. \ref{exemples1} and \ref{exemples2}, display four examples of \rxte/PCA and \inte/JEM-X lightcurves with X-ray dips, for classes $\beta$, $\nu$ and $\lambda$. The corresponding HRs displayed in panels (c) and (f) illustrate the sharp spectral hardening that occurs during dips. The end of each dip is marked by a short spike, particularly visible in $\beta$ and $\nu$ classes. In the radio range, oscillations or isolated flares follow this spike.

\subsection{Consecutive flares}
In order to compare the flare characteristics to the X-ray data, our simple model needed to be refined, notably in the case of consecutive flares. Indeed, several radio lightcurves of \grs\ are characterized by quasi-sinusoidal modulations, which can be interpreted as repeated discrete ejections. In this case, the emission coming from different flares overlaps in the lightcurve. In order to disentangle the emission from each one of them, we modeled these successive ejections using multiple identical flares, separated by a constant interval $t_{lag}$. We checked that, when looking at a sequence of a few radio flares, the time lag between two maxima remains constant within a few \% (Fig. \ref{exemples2}). 

This ``multiflare" approach can be compared to the previous approach used on the 687 flares. By taking the overlaps into account, the refined model leads to values that are lower by at most 10 \% for the amplitude, width and fluence of a given flares. This has to be compared to the uncertainties on these parameters, which lie between $\sim$10--50 \%. Therefore, in the general study above, this effect is not crucial in the determination of the global characteristics of the flares. Since the refined model also leads to lower $\chi_\nu^2$ values, the main source of uncertainty that now remains is the level of the background radio flux.

\begin{figure}
	\plotone{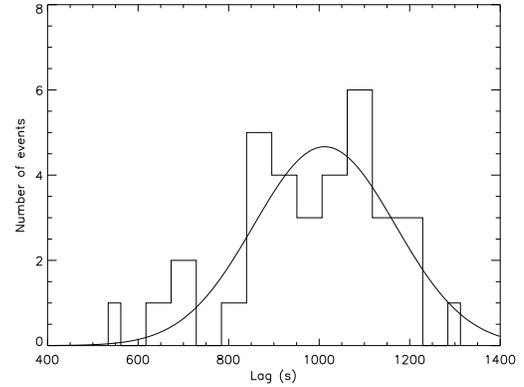}
	\caption{Distribution of the delay between the X-ray spike and the maximum of the subsequent radio flares. The histogram has been computed for the 34 flares for which the spike was visible. The continuous line is a gaussian fit to the histogram.}
	\label{histos_spike}
\end{figure}

\subsection{Confidence indices}
\label{confi}
However, in the case of a strong background, this level is poorly constrained. Indeed, there is a degeneracy between the offset level $K_{0}$ and the other parameters: if $K_{0}$ is fixed to lower values, the fitting routine will change the other parameters to increase the overlap between flares, which will lead to an acceptable fit. This effect is particularly true in the case of strong overlap.

In order to quantify this issue, we attributed a ``confidence index" to our data. This index is based on two criteria: the level of background flux, and the morphology of the observed flares. Indeed, when a radio oscillation is fully shaped, in particular with a clear exponential tail, we have good confidence that the background subtraction is accurate. On the other hand, in the case when only the ``tip" of the flare is visible, the background level cannot be precisely constrained, which is a source of higher uncertainty for the determination of flare properties.

We attributed a confidence index of 1 to the most reliable data: observations with fully shaped flares, either isolated or on top of a low background emission ($\la$15 mJy, Fig. \ref{exemples1}). An index of 2 corresponds to observations where the exponential decrease is still clearly visible, with moderate background emission ($\la$30 mJy, Fig. \ref{exemples2}, left). Finally, observations with almost sinusoidal oscillations on top of a strong background (up to 120 mJy, Fig. \ref{exemples2}, right) correspond to index 3. Thus, parameters deduced from index 3 observations should be considered with care, while index 1 observations would produce reliable parameters.

Among the observations listed in Table \ref{logg}, four observations had to be excluded from our data set: on three observations, two successive flares are visible but not distinguishable, while on one more observation the radio data are too noisy to get any constraint on the flare.

\subsection{X-ray dips and occurrence of radio flares}

To measure the characteristics of the X-ray dips, we used the following definition. Time 0 is defined as the time when the phase of highly variable X-ray flux ends. This time is also the first point of spectral hardening (visible in the HR). We used the position of the maximum flux of the X-ray spike to mark the end of the dip, whose duration will be noted $\Delta t$. At this time, the HR has gone down to values close to the pre-dip phase.

In order to measure the amplitude of the final spike, the difficulty lies in the determination of the beginning of the spike: there is no particular feature marking this point. Instead, we used the minimum flux of the dip, whose value is close to the flux near the end of the dip. Thus, the amplitude was taken to be the difference between the top of the spike and the minimum flux of the dip.

\begin{figure}
	\plotone{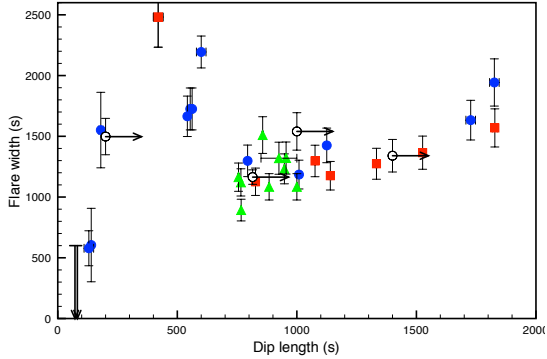}
	\caption{Width of the radio flares ($t_0$), as a function of the length of the preceding Hard X-ray dip $\Delta t$. Blue and black circles correspond to confidence index 1 observations, red squares to index 2, and green triangles to index 3. The error bars are estimated at the $3\sigma$ confidence level (note that errors on index 3 data are questionable).}
	\label{blob}
\end{figure}

Using the parameters given by the fits, we looked for links between the X-ray and radio behaviors of \grs. We confirm the association between X-ray dips and radio flares: among the 54 X-ray dips listed on table 1, 52 are directly followed by a radio flare, only the 2 shortest dips being not followed by a detectable activity. We thus confirm that X-ray dips during $\alpha$, $\beta$, $\lambda$ and $\nu$ classes are always followed by radio flares. Fig. \ref{histos_spike} shows the distribution of the delay between the X-ray spike and the peak of the radio flare, for 34 X-ray dips. For 18 other dips, a flare is also detected but the date of the spike is not known because of the lack of X-ray coverage.

The average delay between the X-ray spike and the peak of the radio flare is 1040 $\pm$ 185 s (at 1$\sigma$), a value consistent with the one calculated by \citet{Rodriguez:2008a}. As for the derived start time of the ejection, according to the van der Laan model, the ejection of matter is coincident with the time of the X-ray spike within less than 5 minutes.

\subsection{Other cyclic observations}
\label{other}
Since this connexion between X-ray dips and radio flares seems ubiquitous, it is interesting to look for cycles \textit{not} followed by detectable flares. Firstly, very short cycles, such as the two $\lambda$ cycles observed on MJD 52171.799, are not followed by a detectable radio activity (upper limit of $\sim$2 mJy). Given that $\lambda$ cycles longer that 100 s, such as the one displayed on Fig. \ref{exemples1}, right, are followed by a weak radio flare, an explanation may simply be that such flares trigger too small ejections to be detected by the Ryle Telescope.

Then, the $\theta$, $\kappa$ and $\rho$ classes are not followed by distinct radio flares, either. These classes were not included in the analysis so far. $\kappa$ and $\rho$ classes display very short cycles, between 10 and 50 s. These cycles recur very quickly, with less than 100s between two consecutive X-ray spikes. Our fits show that the radio flares of \grs\ peak around 1000 s after the end of the cycle, and last typically 1200 s, thus they could not be seen individually during these classes. Note that a weak radio flux is detectable, at $\sim$3--4 mJy (as already seen by \citet{KleinWolt:2002}), which could be compatible with the sum of very small radio ejections occurring after each cycle.

$\theta$ classes display longer cycles, with X-ray dips lasting typically between 300 and 600 s. These cycles also recur quickly, with less than 1000s between two consecutive X-ray spikes. In this case too, individual flares will not be distinguishable, but should produce a varying radio flux. 

\begin{figure}
	\plotone{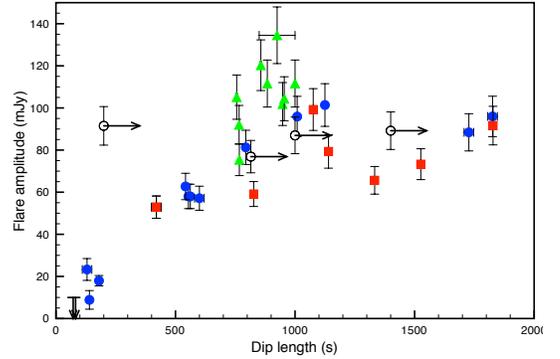}
	\caption{Amplitude of the flare (normalization $S_{max}$), as a function of the dip-length $\Delta t$. The symbols are identical to Fig. \ref{blob}.}
	\label{blob_K}
\end{figure}

Radio observations during this class show a strong radio flux, at $\sim$10--80 mJy \citep{KleinWolt:2002}. This flux is highly variable, but displays no specific pattern. This value of $\sim$10--80 mJy is significantly lower than the maximum fluxes observed after $\beta$ and $\nu$ cycles of comparable duration. Thus, this radio activity could be explained by radio flares, but of lower amplitude. Note that $\theta$ cycles are also characterized by less pronounced X-ray dips: the Hardness Ratio is lower and the minimum flux higher than during other cyclic classes. This may explain their different behavior regarding radio activity.

To put it in a nutshell, observations of $\theta$, $\kappa$ and $\rho$ classes are not entirely conclusive. On the one hand, they can be understood within the framework of radio ejections: the distinct emissions of matter are simply too quick to be separated in the radio lightcurve. On the other hand, the observed radio flux is lower than expected if we followed the trend given by $\beta$, $\lambda$ and $\nu$ classes. This probably means that several features determine the characteristics of the radio activity: the length of the X-ray dips may not be the only parameter required to explain all the radio flares.

\subsection{A correlation between duration of X-ray dips and fluence/amplitude?}
Now, let us look for correlations between the various parameters at hand. The amplitude of the final X-ray spike is not related to the width or amplitude of the following radio flare: the maximum X-ray flux seems to be random. No link was found between the minimum flux during a dip and the characteristics of the following flare either. A slight link, although not statistically significant, was found between the minimum X-ray flux and the duration of the dip: lower minimum fluxes tend to be related to longer dips. This link was already mentioned by \citet{Belloni:1997} in the case of class $\kappa$ observations.

\begin{deluxetable}{cccccc}
\tablewidth{0pt}
\tablecaption{Significance of the correlation between the length of X-ray dips $\Delta t$ and the amplitude $S_{max}$, width $t_0$ and fluence $L_{15GHz}$ of subsequent radio flares. Values are expressed in levels of significance for the two tests.\label{correl}}
\tablehead{\colhead{Quantities} & \colhead{Test}  & \colhead{Index 1} &
\colhead{Index 1 \& 2} & \colhead{All data}
}

\startdata
$L_{15GHz}$ vs. $\Delta t$ & Spearman & $>$99.99\% &          98.4\% &        98.3\% \phn \\
                                   & Pearson     & 87.3\% \phn &          65.1\% &        56.5\% \phn \\
$S_{max}$ vs. $\Delta t$     & Spearman  & 99.98\%       & $>$99.99\% &         99.92\%       \\
                                   & Pearson     & 85.6\% \phn &          80.5\% &        59.0\% \phn \\
$t_0$ vs. $\Delta t$            & Spearman  & 77.0\% \phn & \phn 1.0\%   & \phn 1.0\%   \phn \\
                                   & Pearson     & 39.1\% \phn & \phn  1.2\%   & \phn 2.3\%   \phn \\

\enddata
\end{deluxetable}

\begin{figure}
	\plotone{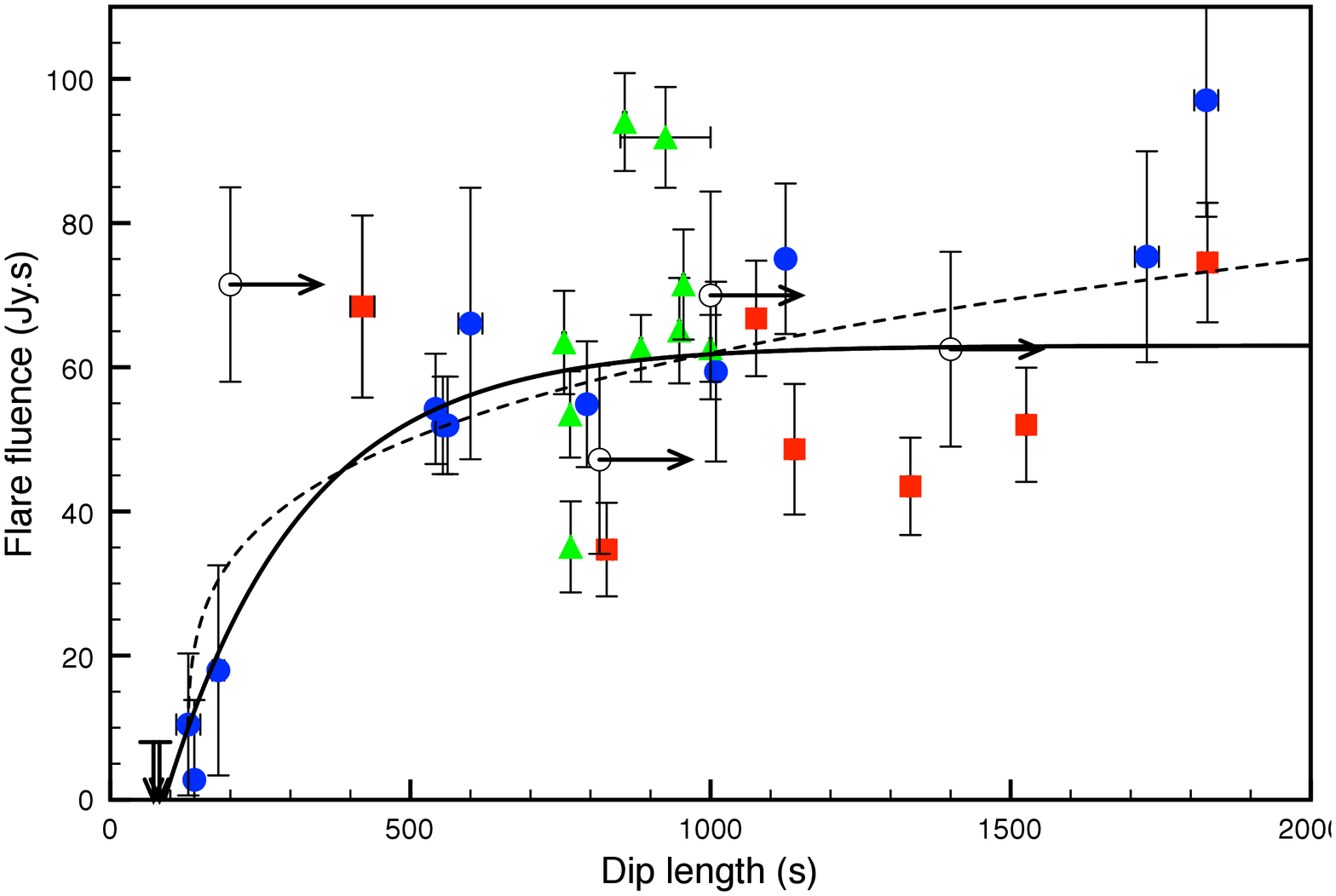}
	\plotone{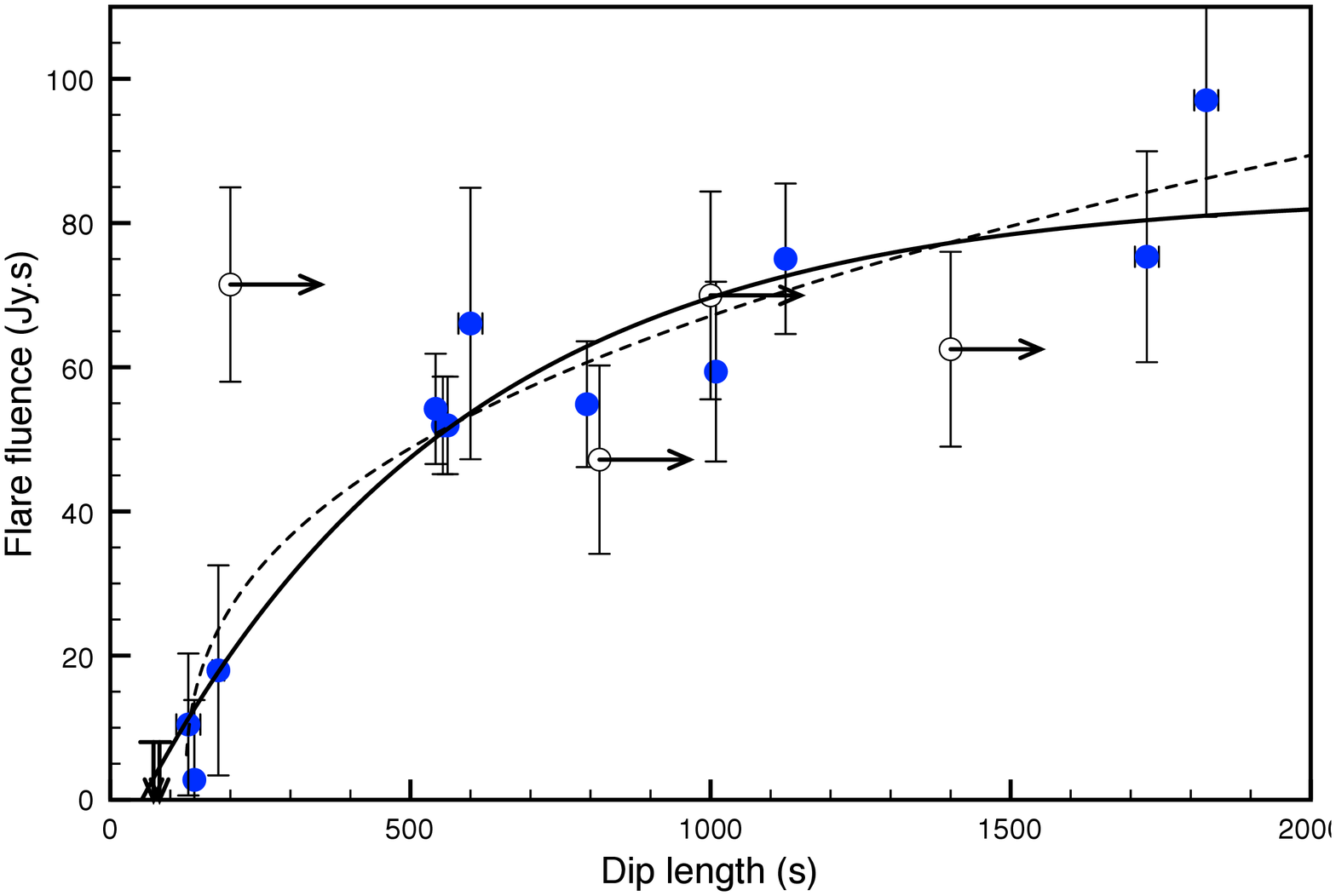}
	\caption{Fluence of the radio flares as a function of the dip-length $\Delta t$. The symbols are identical to Fig. \ref{blob}. The top panel displays all available observations, while the lower one displays index 1 data only. The continuous line is an exponential fit and the dashed line a power-law fit to the data on each panel.}
	\label{resume}
\end{figure}

Finally, a more pronounced trend is visible between the duration of the dips $\Delta t$ and the parameters of the subsequent flares: longer dips seem to be followed by more important radio ejections (Fig. \ref{blob}, \ref{blob_K} and \ref{resume}). In particular, a correlation is visible between $\Delta t$ and the amplitude $S_{max}$ of the following radio flare. A correlation is also visible between $\Delta t$ and the fluence $L_{15GHz}$ of the following radio flare.

To quantify this, we used the Pearson and Spearman tests of correlation. The Spearman test detects any monotonic correlation in the data, while the Pearson test detects linear correlation. Values of these indices are reported in Table \ref{correl} in confidence levels. The two tests show a strong correlation between the length of the X-ray dip $\Delta t$ and the maximum of the following flare $S_{max}$. The correlation between the length of the dips and the fluence $L_{15GHz}$ is also high. Spearman and Pearson tests are not significant in the case of length of the X-ray dip $\Delta t$ versus width $t_0$ of the flares. 

Note, however, that the $L_{15GHz}$ vs. $\Delta t$ and $S_{max}$ vs. $\Delta t$ correlations are less pronounced when ignoring the five points corresponding to $\Delta t < 200 s$. In the case of $S_{max}$ vs. $\Delta t$, the level of significance for the Spearman test drops to values between 60\% and 98\%, for the ``all data" and ``Index 1" subsets, respectively. Indeed, as visible on Fig. \ref{blob_K}, the index 1 data are still correlated when ignoring the $\lambda$ points. In the case of $L_{15GHz}$ vs. $\Delta t$, however, the non-$\lambda$ points are almost compatible with a constant.

During the $\lambda$ observations, there is no radio emission prior to the X-ray spike, so the characteristics of the flare are well constrained. We therefore have good confidence on the robustness of these values. However, a crucial test for the reality of the $S_{max}$ vs. $\Delta t$ and $L_{15GHz}$ vs. $\Delta t$ correlations would be to observe X-ray dips in the range 200--500 s. More observations are needed to populate this range in order to refine the link between X-ray dips and radio flares.

\medskip
Looking at the fluence vs. dip-length relationship, the correlation does not seem to be linear nor affine. Indeed, short X-ray dips produce much weaker radio flares than long dips (see Fig. \ref{resume}). This is consistent with the statement of \citet{KleinWolt:2002}, who mentioned that dips shorter than 100s do not seem to be followed by radio flares.

Therefore, we looked for other functions to describe the relationship between the (radio) fluence $L_{\mathrm{15}Ê\textrm{ } \mathrm{GHz}}$ and the length $\Delta t$ of the X-ray dip than a simple linear or affine law (Fig. \ref{resume}). Two types of functions produced good results, which are listed in Table \ref{expo}. Firstly, the data are well described by a function of the form $L_{\mathrm{15}Ê\textrm{ } \mathrm{GHz}}(\Delta t) = A (\Delta t-\Delta t_0)^\gamma$, with $\gamma$ between 0.17 and 0.35 depending on the subset considered. Then, an exponential function of the form $L_{\mathrm{15}Ê\textrm{ } \mathrm{GHz}}(\Delta t) = L_{max} (1-e^{-\frac{\Delta t-\Delta t_{min}}{\tau}})$ provided very similar $\chi_\nu^2$, with a characteristic time $\tau$ between 100 and 500 s, and a saturating fluence $L_{max}$ between 55 and 80 Jy.s.

Both models are truncated at short dip-length, as visible on the plots. In these two cases, the quantities $\Delta t_0$ and $\Delta t_{min}$ correspond to the length of the X-ray dip below which no flare is detected ($L_{\mathrm{15}Ê\textrm{ } \mathrm{GHz}}=0$). They measure $\Delta t_0$ and $\Delta t_{min}$ between 50 and 130 s. This is compatible with the empirical observation that no flare are detected after dips shorter than $\sim$100 s.

\section{Discussion}
\label{Discut}
Our analysis extends the previous results on the occurrence of radio flares. \citet{KleinWolt:2002} suggested that, during $\beta$ and $\nu$ classes, long and spectrally hard State C intervals are followed by radio oscillations. \citet{Rodriguez:2008a} extended this connexion to $\lambda$ classes, and identified the X-ray spike at the end of the X-ray dips as the trigger to the ejection. With an extended data set, we confirm that, during $\alpha$, $\beta$, $\lambda$ and $\nu$ classes, every X-ray dip in the lightcurve is followed by a radio flare. Moreover, the beginning of this flare coincides with the X-ray spike within less than 300s. On $\kappa$ and $\rho$ classes, we lack the sensitivity to confirm or reject a similar behavior. Finally, $\theta$ classes may also follow this scheme with, however, radio flares significantly smaller than during $\beta$ and $\nu$ dips of comparable duration.

\begin{deluxetable*}{ccccccccc}
\tablewidth{0pt}
\tablecaption{Values obtained by fitting power-law and exponential functions to the dip length vs fluence data. The 2 functions used were of the form $L_{\mathrm{15}Ê\textrm{ } \mathrm{GHz}}(\Delta t) = L_{max} (1-e^{-\frac{\Delta t-\Delta t_{min}}{\tau}})$ and $L_{\mathrm{15}Ê\textrm{ } \mathrm{GHz}}(\Delta t) = A (\Delta t-\Delta t_0)^\gamma$. Errors are given at the 68\% confidence level (1$\sigma$). \label{expo}}
\tablehead{\colhead{Set} & \colhead{$L_{max}$}  & \colhead{$\tau$} &
\colhead{$\Delta t_{min}$} & \colhead{$\chi^2_\nu$} & 
\colhead{$A$}  & \colhead{$\Delta t_0$} & \colhead{$\gamma$} & \colhead{$\chi^2_\nu$}\\
\colhead{} & \colhead{(Jy.s)}  & \colhead{(s)} &
\colhead{(s)} & \colhead{} & 
\colhead{(Jy.s)} & \colhead{(s)}  & \colhead{} & \colhead{}
}

\startdata
Index 1        & 80$\pm$13 &           490$\pm$190       & \phn 50$\pm$70 & 0.34  & \phn 6$\pm$4 & 125$\pm$15        & 0.35$\pm$0.10 & 0.29 \\
Index 1 \& 2 & 55$\pm$20 & \phn  90$\pm$67 \phn    &        120$\pm$20 & 2.09 &        77$\pm$6 & 130$\pm$2 \phn & 0.17$\pm$0.06 & 2.11  \\
All data        & 63$\pm$10 &          220$\pm$75 \phn  & \phn 90$\pm$40 & 3.90  & \phn 9$\pm$2 & 129$\pm$6 \phn & 0.28$\pm$0.02 & 4.08 \\

\enddata
\end{deluxetable*}

Using the available $\beta$, $\lambda$ and $\nu$ observations, a trend is visible between the length of the X-ray dips and the characteristics of the following flare: the longer the X-ray dip is, the bigger the following radio flare will be. This result does not directly depend on the plasmon model used to characterize the radio data. Indeed, this model we chose was used mainly to distinguish between the energy emitted during each individual flare in a sequence of radio flares, rather than as a precise description of each flare. Therefore, every ``realistic" function with a fast rise and slow decay able to fit the shape of the radio flares would lead to the same result. In particular, emission from a conical jet or a shock-heated compact jet would have almost the same temporal shape, and thus lead to similar results concerning the basic parameters explored here.

\medskip

This link seems to be related to the length of hard X-ray dips: longer dips lead to more energetic ejections. Although the data are scarce, this link does not seem to be linear, and can be described, for instance, by a power-law or exponential relationship. Given the high uncertainties on the data, it is not possible to discriminate between these functions.

Yet, in the second case, one could think of an attractive scenario to explain the relationship. Observations show that the beginning of the dip is marked by a quick increase in the accretion disk inner radius. Then, during the dip itself, the disk draws closer to the BH \citep{Belloni:1997, Migliari:2003, Rodriguez:2008b}. This evolution brings more gravitational energy close to the BH. Let us suppose that, during the Hard X-ray dip, energy is somehow extracted from the accretion disk, and accumulated in the surrounding medium. Thus, the luminosity of the disk and corona slowly increase. At the same time, the amount of energy lost by the corona in a given time grows as the energy density of the corona increases. Thus, the total energy stored inside the corona saturates after, say, $\sim$1000s. Then, at some point, this energy is released in the form of a quickly expanding blob of matter. In the X-ray lightcurve, this ejection is marked by a short X-ray spike \citep{Mirabel:1998, Rodriguez:2008a}. Note that the ejected material can come from the corona itself, or from the inner accreting disk; either way, after the spike the corona is not visible anymore, and the disk is closer to the BH.

\medskip

In the case of a jet (either steady or shock-heated), the scenario is similar: during the dips an injection of material into the jet from the corona takes place \citep{KleinWolt:2002}. Assuming a more or less constant injection rate, longer X-ray dips will thus be radio-brighter, as more material is ejected. Furthermore, during the dip, as the disk moves back in towards the compact object, one can assume that it reduces the injection rate, which would explain the apparent saturation in the flare fluence. Then, at some point, the disk is so close to the BH that it suppresses the injection of particles. The X-ray dip ends, and radio emission is detected afterwards.

However, the jet interpretation suffers from an important caveat: it predicts lightcurves different from the observed radio flares. Indeed, using the conical jet from \citet{Hjellming:1988}, it is possible to model the radio emission from a transient jet. During a given dip, the jet would be active for $\sim$30 min. With a velocity of about $v_{exp}=0.8c$, the jet would have a total extension of $\sim4.10^{13}$cm, or about $\sim10^7 R_S$. The resulting synchrotron emission would thus emanate from electrons at very different temperatures, due to adiabatic expansion. When looking at a fixed radio frequency, the lightcurve would then be broader than that of a single plasmon. Besides, the radio emission would be detected from the beginning of the X-ray dip, or with a constant time-lag with it. Since the data show that the radio emission begins within a few minutes from the X-ray spike at the end of the dip, with no dependance on the length of the dip, this interpretation seems less probable than the discrete ejection model.

\medskip
It is interesting to note that the distribution of width of the radio flares is quite peaked while that of the amplitude not. In the framework of the plasmon model \citep{vanderLaan:1966}, the width of the flare is related to the initial physical size of the plasmon. Thus, the relative sharpness of this distribution means that the initial radius of the plasmon is roughly always the same. It is tempting to interpret that this radius may be constrained by the inner radius of the disk. On the other hand, the maximum amplitude of the flare is related to the initial temperature and energy distribution of the ejected plasma. Thus, the relative broadness of the amplitude distribution could mean that the amount of energy stored inside the plasma prior to the ejection is more variable.

If we assume that the amount of energy stored inside the plasma depends on the duration of the preceding X-ray dip, then this relative broadness can be easily explained. Indeed, during shorter dips the input of energy would be shorter, thus the amplitude of the following flare would be smaller. Since the duration of the X-ray dips is highly variable, the amplitude of the flares will also be variable. On the other hand, the medium which stores this energy will always be located in the same region, close to the inner radius of the disk; if its physical size is constrained by the disk, then the width of the flares will be roughly constant.

One possible interesting model which could provide a more precise interpretation of this behavior is the ``Magnetic Flood" model \citep{Tagger:2004}. This model was proposed to account for the behavior of \grs\ during the $\beta$ class. It relies on the Accretion-Ejection Instability \citep{Tagger:1999}, which would develop during the X-ray dip. This instability relies on the presence of a moderate magnetic field in the inner parts of the disk, and can produce Low-Frequency Quasi-Periodic Oscillations (LFQPOs), as well as feed a corona with matter from the disk. Then, if the magnetic configuration is favorable, a sudden reconnection event can occur between magnetic fields of opposite polarities in the accretion disk. This reconnection event would produce the spike seen in X-rays, and power the ejection of matter.

\begin{acknowledgments}
The authors thank the referee for the thorough and thoughtful review that led to significant improvement in the paper. JR acknowlegdes partial funding from the European Community's Seventh Framework Programme (FP7/2007-2013) under grant agreement number ITN 215212 ``Black Hole Universe". This work has been (partly) financially supported by the GdR PCHE in France. This research has made use of data of the RXTE satellite operated by NASA, obtained through the High Energy Astrophysics Science Archive Center Online Service, provided by the NASA/Goddard Space Flight Center. Based on observations with \inte, an ESA project with instruments and science data centre funded by ESA member states (especially the PI countries: Denmark, France, Germany, Italy, Switzerland, Spain), Czech Republic and Poland, and with the participation of Russia and the USA. The Ryle Telescope was operated by the University of Cambridge and supported by STFC.
\end{acknowledgments}


\end{document}